\documentclass[twocolumn,preprintnumbers,floats,prd,amssymb,floatfix,nofootinbib,balancelastpage,superscriptaddress,amsmath]{revtex4-1}

\pdfoutput=1
\usepackage[utf8]{inputenc}
\usepackage{amssymb}
\usepackage{amsmath}
\usepackage{amsfonts}
\usepackage{graphicx}
\usepackage{color}
\usepackage{xspace}

\usepackage[normalem]{ulem} 

 \usepackage[section]{placeins}
\usepackage{afterpage}

\usepackage{float}
\usepackage{slashed}

\usepackage{multirow,rotating}
\usepackage{hyperref}

\def\gsim{\raise0.3ex\hbox{$\;>$\kern-0.75em\raise-1.1ex\hbox{$\sim\;$}}}
\def\lsim{\raise0.3ex\hbox{$\;<$\kern-0.75em\raise-1.1ex\hbox{$\sim\;$}}}

\newcommand{\ba}[1]{\begin{eqnarray} \label{(#1)}}
\newcommand{\ea}{\end{eqnarray}}

\newcommand{\lam}{\lambda}

\definecolor{dcolour}{rgb}{.5, .5, .5}

\def\gsim{\raise0.3ex\hbox{$\;>$\kern-0.75em\raise-1.1ex\hbox{$\sim\;$}}}
\def\lsim{\raise0.3ex\hbox{$\;<$\kern-0.75em\raise-1.1ex\hbox{$\sim\;$}}}

\newcommand{\neutralino}[1]{\tilde{\chi}_{#1}^0}

\newcommand{\iab}{\rm ab^{-1}}
\newcommand{\ifb}{\rm fb^{-1}}

\begin{document}

\preprint{\parbox[t]{3.3cm}{BONN-TH-2019-01}}

\title{Long-lived light neutralinos at future $Z-$factories}

\author{Zeren Simon Wang}
\email{zerensimon.wang@apctp.org \,(Corresponding author)}
\affiliation{Asia Pacific Center for Theoretical Physics (APCTP) - Headquarters San 31,\\ Hyoja-dong, Nam-gu, Pohang 790-784, Korea}
\affiliation{Physikalisches Institut der Universit\"at Bonn, Bethe Center for Theoretical Physics, \\ Nu{\ss}allee 12, 53115 Bonn, Germany}

\author{Kechen Wang}
\email{kechen.wang@whut.edu.cn}
\affiliation{Department of Physics, School of Science, Wuhan University of Technology,\\
430070 Wuhan, Hubei, China}

\begin{abstract}
Future lepton colliders such as the CEPC and FCC-ee would run as high-luminosity $Z-$boson factories, which offer a unique opportunity to study long-lived particles which couple to Z-bosons.
In order to exemplify this particular advantage, in this work 
we consider one benchmark physics scenario where the long-lived lightest neutralinos pair $(\tilde{\chi}_1^0\tilde{\chi}_1^0)$ is produced from $Z-$decays in the context of the R-parity violating supersymmetry.
Our analysis indicates that when assuming BR$(Z\rightarrow \tilde{\chi}_1^0\tilde{\chi}_1^0) = 10^{-3}$ and $m_{\tilde{\chi}_1^0} \sim 40$ GeV, the model parameter $\lambda'_{112} / m^2_{\tilde{f}}$ can be discovered down to as low as $\sim 1.5 \times 10^{-14}$ ($3.9 \times 10^{-14}$) GeV$^{-2}$ at the FCC-ee (CEPC)  with center-of-mass energy $\sqrt{s} = 91.2$ GeV and 150 (16) $\iab$ integrated luminosity.
These limits exceed the projected sensitivity reaches of the ATLAS experiment at the HL-LHC and the proposed LHC experiments with far detectors (AL3X, CODEX-b, FASER, and MATHUSLA).
\end{abstract}
\keywords{}


\vskip10mm

\maketitle
\flushbottom
%
%

\section{Introduction}
\label{sec:intro}
Long-lived particles (LLPs) arise in many physics scenarios beyond the Standard Model (BSM) and are often motivated by dark matter or the massive neutrinos.
While at the Large Hadron Collider (LHC), efforts have been mostly focused on searching for prompt decays of new heavy particles, it is also legitimate to look for exotic signatures of displaced vertices stemming from LLPs.
For reviews, see, e.g., Refs.~\cite{Alimena:2019zri,Lee:2018pag}.

Supersymmetry \cite{Nilles:1983ge,Martin:1997ns} (SUSY) has been one of the leading candidates of BSM physics since it offers elegant solutions to many important fundamental physics problems such as the hierarchy problem \cite{Gildener:1976ai,Veltman:1980mj}.
The mass eigenstates of electroweak gauginos predicted by SUSY models are known as neutralinos and charginos.
While lower mass bounds on charginos have been derived from the LEP data~\cite{Patrignani:2016xqp}, the limits on the mass of the lightest neutralino are much looser.
If the GUT-motivated relation between the gaugino mass terms $M_1$ and $M_2$, $M_1 \approx 1/2\, M_2$, is not imposed and the dark matter in the universe does not comprise of the lightest neutralino, $\mathcal{O}(10)$ GeV-scale and even massless neutralinos, which are necessarily bino-like, are still allowed by experimental and observational data \cite{Choudhury:1999tn,Hooper:2002nq,Bottino:2002ry,Dreiner:2009ic,Vasquez:2010ru,Calibbi:2013poa,Dreiner:2003wh,Dreiner:2013tja}, though they must decay with a lifetime much shorter than the age of the universe so as to be consistent with the dark mater density.

R-parity-violating SUSY (RPV-SUSY) (see Refs.~\cite{Barbier:2004ez,Dreiner:1997uz,Mohapatra:2015fua} for reviews) naturally leads to decays of the lightest neutralino via RPV couplings, allowing for light neutralinos of $\mathcal{O}$(GeV) mass.
The smallness of the neutralino mass and RPV couplings renders the lightest neutralino long-lived, potentially resulting in displaced vertex signatures at colliders. 
Such signatures may be observed at a variety of experiments including the fixed target experiment SHiP \cite{Anelli:2015pba}, the LHC experiment ATLAS \cite{Aad:2008zzm} or some proposed future detectors: CODEX-b \cite{Gligorov:2017nwh}, MATHUSLA \cite{Curtin:2018mvb}, FASER \cite{Feng:2017uoz}, and AL3X \cite{Gligorov:2018vkc}.
Studies of the light neutralinos as LLPs decaying via RPV couplings in these experiments have been performed in Refs.~\cite{Helo:2018qej,Dercks:2018eua,Dercks:2018wum,deVries:2015mfw,Liu:2015bma}.
In these references, two production mechanisms of the lightest neutralino have been taken into account: 1) single production from rare $B-$ and $D-$meson decays via RPV couplings, 2) pair production from rare $Z-$boson decays via the Higgsino components.
In this study, we focus on the latter in the context of future lepton colliders running at the $Z-$pole mode.

While the LHC is planned to be upgraded to high-luminosity LHC (HL-LHC) in the coming years, several next-generation new colliders have been proposed and are under development.
Among them are the Circular Electron Positron Collider (CEPC)~\cite{An:2018dwb,CEPCStudyGroup:2018ghi} to be built in China and the Future Circular Collider (FCC)~\cite{Abada:2019lih} at CERN as the successor of the LHC.
The FCC would, as currently planned, start with an electron-positron collision mode, known as the FCC-ee \cite{Abada:2019zxq}.

Both the CEPC and FCC-ee would operate at the $Z-$pole mode (with center-of-mass energy $\sqrt{s} = 91.2$ GeV) for 2-4 years, 
making them excel in search for $Z-$coupled LLPs for various reasons.
Firstly, the CEPC and FCC-ee running at the $Z-$pole mode are expected to produce a terascale number of $Z-$bosons, exceeding the HL-LHC by approximately one order of magnitude and LEP by $\sim$ five orders of magnitude.
As LLPs are usually very feebly coupled to standard model (SM) particles, their production cross sections at colliders are tiny.
Such a large number of $Z-$bosons produced at the CEPC and FCC-ee could therefore significantly enhance the discovery sensitivities of LLPs produced from rare $Z-$boson decays.
Secondly, given the not so large center-of-mass energy at $\sqrt{s} =m_Z$, the SM long-lived hadrons arising from $Z-$decays are featured with a much shorter boosted decay length, compared to those at a hadron collider such as the LHC.
This leads to a much smaller number of SM background events for LLP searches.
Moreover, 
since the momenta of the incoming electrons and positrons are known at lepton colliders, it is possible to reconstruct the momenta of all the final state particles, allowing missing energy measurement and more stringent cuts for the signal.
Finally, the detector of the CEPC or FCC-ee is designed to employ comparable angular coverage and volume as the ATLAS/CMS detector. Its solid angle coverage is larger than the proposed far detectors at the LHC as well.

Studies have investigated the discovery potential of future lepton colliders for a variety of new physics and SM scenarios related to $Z-$properties \cite{Blondel:2014bra,Abada:2014cca,Lesiak:2018vdf,Monteil:2016pjz,Liu:2017zdh,Ding:2019tqq,Chang:2018pjp,Carena:2003aj,Ke:2009sk,Jin:2010wg,Huang:2014cxa,Xu:2014ova,Durieux:2015hsi,Cao:2010na,Wang:2011qz,Domingo:2011uf,Wang:2011zzt,Ghosh:2014rha,Flacke:2016szy,Blinov:2017dtk,Gao:2017tgx,Yu:2014ula,Fabbrichesi:2017zsc}.
In this work, we study the discovery potential of the CEPC and FCC-ee when searching for LLPs produced from $Z-$boson decays, by considering one particular benchmark scenario: the rare decays of the $Z-$bosons into a pair of long-lived lightest neutralinos ($Z\rightarrow \tilde{\chi}_1^0\tilde{\chi}_1^0$) in the RPV-SUSY.
As the official parameters for the $Z-$pole running mode are not released yet, another two proposed future $e^+ e^-$ colliders, ILC (International Linear Collider) \cite{Fujii:2017vwa} and CLIC (Compact Linear Collider) \cite{deBlas:2018mhx} are not considered in this study.

This paper is organized as follows.
We explain the physics accounting for the decay processes: $Z\rightarrow \tilde{\chi}_1^0\tilde{\chi}_1^0$ and neutralino decays in Sec.~\ref{sec:theory}.
In Sec.~\ref{sec:detector-and-simulation}, we introduce the fiducial volume of detectors at future $Z-$boson factories, and present the simulation procedure.
The possible trigger setup and SM background issues are also discussed in this section.
In Sec.~\ref{sec:results} we show our numerical results and compare the sensitivity reaches at the CEPC and FCC-ee with those at the current and proposed LHC experiments.
We summarize the paper in Sec.~\ref{sec:conclusions}.

\section{Pair Production of light Neutralinos and RPV-Supersymmetry}
\label{sec:theory}
In this section, we explain the production and the decay mechanisms of the lightest neutralino which are considered in this paper.
The lightest neutralino can be produced in a variety of physics processes.
In this paper, we focus on their pair production from on-shell $Z-$boson decays, taking advantage of the large $Z-$boson production at the future high-luminosity lepton colliders.
A $Z-$boson is coupled to two lightest neutralinos via the Higgsino components, leading to its decay to a pair of neutralinos, if $m_{\tilde{\chi}_1^0}<m_Z/2$.
While light neutralinos are necessarily bino-like and include only small Higgsino components, the sufficiently copious production of the $Z-$bosons may still compensate for it.
In Ref.~\cite{Helo:2018qej}, it is discussed that the current lower limit on the Higgsino parameter $\mu$ in the supersymmetry models, obtained in LEP \cite{Patrignani:2016xqp} and ATLAS \cite{Aaboud:2017leg} experiments, points to a calculated branching ratio BR($Z\rightarrow \tilde{\chi}_1^0\tilde{\chi}_1^0$) just below the experimental upper limit $\sim 0.1\%$ which is derived from the invisible width of the $Z-$boson measured at LEP \cite{Patrignani:2016xqp}\footnote{There is no tension with respect to the experimental bound on the Higgs invisible width \cite{Helo:2018qej}, either.}.
In this study, we treat BR($Z\rightarrow \tilde{\chi}_1^0\tilde{\chi}_1^0$) hence as an independent parameter, disregarding the SUSY parameters affecting  $\Gamma(Z\rightarrow \tilde{\chi}_1^0\tilde{\chi}_1^0)$.

In the Minimal Supersymmetric Standard Model (MSSM) \cite{Nilles:1983ge,Haber:1984rc}, an implicit ingredient is R-parity.
R-parity conservation renders the lightest neutralino stable if it is the lightest supersymmetric particle (LSP) and it serves as a cold DM candidate.
However, it is equally legitimate to consider the R-parity violating MSSM (RPV-MSSM) \cite{Allanach:2003eb} and study its implications in collider searches.
With R-parity broken, the lightest neutralino decays to SM particles and cannot be considered as a DM candidate.
In this paper, we assume R-parity violation and investigate the potential of detecting the lightest neutralino of $\mathcal{O}$(1-10 GeV) mass via its decay products.
Since we will consider neutralinos decay to a kaon, we do not study neutralinos of mass below the kaon mass $\sim 500$ MeV.
The RPV part of the full superpotential in the RPV-MSSM, $W_{\text{RPV}}$, can be written as:
\begin{eqnarray}
W_{\text{RPV}}&=&\mu_i H_u \cdot L_i + \frac{1}{2}\lambda_{ijk}L_i \cdot L_j \bar{E}_k \nonumber\\
&& + \lambda'_{ijk} L_i \cdot Q_j \bar{D}_k + \frac{1}{2}\lambda''_{ijk}\bar{U}_i \bar{D}_j \bar{D}_k,
\end{eqnarray}
where the first three sets of operators violate lepton number and the last set of operators violate baryon number.
Allowing all these terms to be nonvanishing would lead to a dangerous proton decay rate.
Therefore, one may instead impose certain discrete symmetries, forbidding a subset of all terms and avoiding thus the proton decay rate problem \cite{Dreiner:2005rd,Dreiner:2007vp,Chamseddine:1995gb,Dreiner:2013ala}.
In this study, we focus on the $\lambda' L\cdot Q \bar D$ operators.
For $m_{\tilde{\chi}_1^0}<m_Z/2$ and small $\lambda'$ couplings, the lightest neutralino becomes long-lived and decays after having travelled a macroscopic distance.

\section{Simulation and Detectors}
\label{sec:detector-and-simulation}

In this section, we describe our simulation procedure and introduce the detector setups.
The FCC-ee is planned to run at the $Z-$pole mode for a total of 4 years with the physics goal of 150 $\iab$ integrated luminosity with 2 interaction points (IPs), which would produce in total $5 \times 10^{12}$ $Z-$bosons \cite{Abada:2019zxq}. 
The 10-year operation plan of the CEPC includes two years of $Z-$pole period, expected to generate a total of 16 $\iab$ integrated luminosity with 2 IPs, projected to produce $7 \times 10^{11}$ $Z-$bosons \cite{CEPCStudyGroup:2018rmc}.
We express thus the total number of neutralinos produced as follows:
\begin{eqnarray}
N_{\tilde{\chi}_1^0} = 2\, N_Z \cdot \text{BR}(Z\rightarrow \tilde{\chi}_1^0 \tilde{\chi}_1^0),
\label{eqn:neu1number}
\end{eqnarray}
where $N_Z$ denotes the total number of produced $Z$-bosons and a factor of 2 accounts for the fact that each $Z$-boson decays to two neutralinos.
In order to determine the average decay probability of the neutralinos in the fiducial volume of the detectors, we make use of the Monte-Carlo (MC) simulation tool Pythia 8.205 \cite{Sjostrand:2006za,Sjostrand:2014zea}. 
Since the $Z-$boson is hard-coded to have the SM properties in Pythia, it cannot easily be set to decay into new particles with a certain branching ratio. Pythia is then implemented with the module “New-Gauge-Boson Processes” which allows to generate pure $Z'-$bosons from electron-positron scattering. 
By tuning the $Z'-$boson to have the same mass and couplings to both electrons and neutralinos as the SM $Z-$boson, we are able to extract the kinematics of the processes $e^+ e^- \rightarrow Z$, $Z \rightarrow \tilde{\chi}_1^0 \tilde{\chi}_1^0$ after simulating $10$ thousand events for each point in the parameter space.
The average decay probability in the fiducial volume can then be calculated as
\begin{eqnarray}
\left\langle P[\tilde{\chi}_1^0 \text{ in f.v.}]\right\rangle = \frac{1}{N^{\text{MC}}_{\tilde{\chi}_1^0}} \sum_{i=1}^{N^{\text{MC}}_{\tilde{\chi}_1^0}} P[(\tilde{\chi}_1^0)_i\text{ in f.v.}],
\label{eqn:avedecprob}
\end{eqnarray}
where ``f.v.'' stands for ``fiducial volume'' and $N^{\text{MC}}_{\tilde{\chi}_1^0}$ is the total number of MC-simulated neutralinos.
The computation of the individual decay probability $P[(\tilde{\chi}_1^0)_i\text{ in f.v.}]$ depends on the detector geometries and will hence be detailed later when we introduce the detector setups.
We proceed to write the observed decays of the neutralinos in the fiducial volume as
\begin{eqnarray}
N_{\tilde{\chi}_1^0}^{\text{obs}} = N_{\tilde{\chi}_1^0} \cdot \left\langle P[\tilde{\chi}_1^0 \text{ in f.v.}]\right\rangle \cdot \text{BR}(\tilde{\chi}_1^0 \rightarrow \text{ final state}),\,\,\,\,\,\,\,\,\,\,\,\,
\label{eqn:obsneu1number}
\end{eqnarray}
where $\text{BR}(\tilde{\chi}_1^0 \rightarrow \text{ final state})$ is the branching ratio of the $\tilde{\chi}_1^0$ decays to the final states that we consider.

For calculating the individual decay probability, i.e. $P[(\tilde{\chi}_1^0)_i\text{ in f.v.}]$, we need to take into account the detector setups.
The CEPC is equipped with a baseline detector concept \cite{CEPCStudyGroup:2018ghi}.
In its inner region, there are a silicon pixel vertex detector, a silicon inner tracker, and a Time Projection Chamber which reconstructs the tracks of objects.
For the FCC-ee, two detector designs have been proposed, namely the ``CLIC-Like Detector'' (CLD) \cite{AlipourTehrani:2254048} and the ``International Detector for Electron-positron Accelerators'' (IDEA)\footnote{The CEPC also takes IDEA as an alternative detector concept. \cite{CEPCStudyGroup:2018ghi}} \cite{Abada:2019zxq}.
As the name says, the CLD design is modified from the CLIC detector after taking into account the FCC-ee specificities.
Both detector designs of the FCC-ee employ a setup similar to that of the CEPC baseline detector.

In this paper, we consider the fiducial volume of the detectors as the inner detector consisting of the vertex detector and the tracker.
This choice is conservative and ensures that the electrons produced from the neutralino decays could be reconstructed. 
The signal events considered in this study require at least one neutralino decaying inside the inner detector, while the other could decay either inside or outside the inner detector.
A possible concern is how to efficiently trigger on this signal at future lepton colliders.
We illustrate the trigger possibilities below by considering the case of switching on only one $L Q \bar{D}$ operator, i.e. $\lambda'_{112} L_1 Q_1 \bar{D}_2$.
The lightest neutralino may decay to either charged or neutral particles.
Details about decay modes can be found in Sec.~\ref{sec:results}.

For the events with both neutralinos decaying inside the inner detector, the triggers of our signal could rely on the visible products from both neutralino decays.
The two plots in Fig.~\ref{fig:trigger_distribution_1} show the kinematic distributions of  decay products from two neutralinos for three representative values of $m_{\tilde{\chi}_1^0} = $ 1, 10, and 40 GeV.
$p_T^{\rm sum}(\rm Visible)$ denotes the sum of transverse momentum $p_T$ of all visible decay products, and is centered near the $Z-$boson mass $m_Z$ (except for $m_{\tilde{\chi}_1^0}$ close to $m_Z/2$).
$H_T$ is the sum of $p_T$ of all hadronic decay products, and has a relatively flat distribution peaked around $m_Z/2$ across the kinematically allowed neutralino mass range.
Applying conservative triggers on these observables according to existing LHC searches can be reasonable.
For example, one may estimate the trigger efficiencies at lepton colliders based on the ATLAS tau trigger~\cite{Nedden:2017bvb} which only requires the jet $p_T \gsim 30$ GeV.
The distributions in Fig.~\ref{fig:trigger_distribution_1} indicate that similar triggers at lepton colliders would filter away only a small portion of signal events.

For the events with only one neutralino decaying inside the inner detector, the other neutralino most likely does not deposit any energy in the calorimeters and acts hence as missing energy.
Since at lepton colliders $Z-$bosons are produced at rest, one can infer the missing energy directly as the difference between $m_Z$ and the energy of all visible products of the decaying neutralino.
In fact, the distribution of the missing energy is expected to peak around $m_Z/2$.
Consequently, one trigger on the missing energy with a threshold below $m_Z/2$ should be able to preserve almost all signal events.
Besides, one can still consider a trigger on the $p_T^{\rm sum}(\rm Visible)$, which denotes now the sum of $p_T$ of all visible decay products from the decaying neutralino.
Fig.~\ref{fig:trigger_distribution_2} shows the corresponding distribution for three representative values of $m_{\tilde{\chi}_1^0}=1,$ 10, and 40 GeV.
Following the similar discussion in the previous paragraph,  a trigger requiring $p_T>30$ GeV would remain a large portion of signal events.

Furthermore, even though in this work we are restricted to the inner detector as the fiducial volume, other detector components such as the hadronic calorimeter (HCAL) and the muon spectrometer (MS) can still be employed for LLP searches.
For example, when a LLP, e.g., the lightest neutralino, decays into one electron and two jets inside the HCAL, the latter can see an energy deposit and measure the energy.
Such effect has been exploited at ATLAS to search for displaced lepton jets \cite{ATLAS:2016jza}.
Similarly, charged tracks are also visible in the first two layers of the MS.
Therefore, in principle, even if the LLPs decay outside the inner detector, it is still possible to observe the decay products.
Utilizing these effects lead hence to more trigger possibilities, such as the ratio between energies deposited in the hadronic and electromagnetic calorimeters.

\begin{figure}[t]
\centering
 \includegraphics[width=\columnwidth]{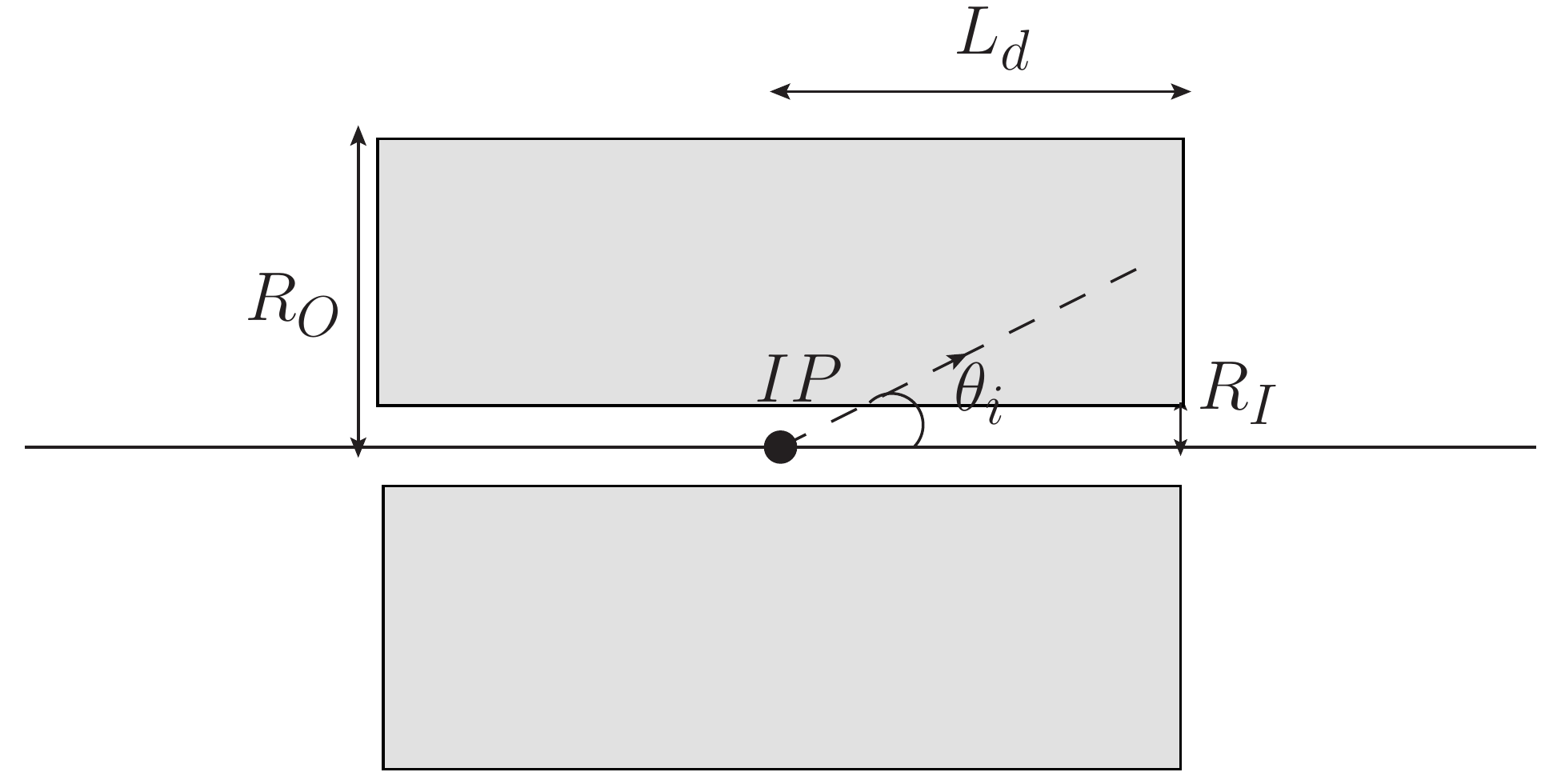}
\caption{General side-view sketch of the fiducial volumes of detector designs for the CEPC and the FCC-ee, with definition of distances and angles used in text. The detectors are cylindrically symmetric around the beam axis. IP denotes the interaction point at the CEPC or the FCC-ee. The dashed line depicts an example neutralino track, with polar angle $\theta_i$.}
\label{fig:detector-sketch}
\end{figure}

Since the designs of all of CEPC baseline detector, FCC-ee CLD, and IDEA detectors are cylindrically symmetric around the beam axis with the IP at the center, we show in Fig.~\ref{fig:detector-sketch} a general side-view sketch of the detector fiducial volume, where $R_I$ is the inner radius of the vertex detector, and $R_O$ and $L_d$ are respectively the outer radius and the half length of the tracker.
Although the various detectors share the same topology, they are designed with different geometrical parameters ($R_I$, $R_O$, $L_d$) and they also have different integrated luminosities of $Z-$boson production as discussed above.
We summarize the relevant information in Table.~\ref{tab:detectors}.

\begin{table}
\begin{center}
\begin{tabular}{c|c|c|c|c}
\hline
\hline
Detector    & $R_I$ [mm] & $R_O$ [m] & $L_d$ [m] & $N_Z$                               \\
\hline
CEPC        & 16       & 1.8    & 2.35    & $7 \times 10^{11}$                  \\
\hline
FCC-ee CLD  & 17       &  2.1   &  2.2  & \multirow{2}{*}{$5 \times 10^{12}$} \\
FCC-ee IDEA & 17       &   2.0   &  2.0   &      \\
\hline
\hline                              
\end{tabular}
\caption{Summary of parameters of the fiducial volume of each detector. $N_Z$ is the total number of $Z-$bosons expected to be produced. The parameters of the CEPC baseline detector are extracted from Refs.~\cite{CEPCStudyGroup:2018ghi,CEPCStudyGroup:2018rmc} 
while the geometries of the CLD and the IDEA detectors of the FCC-ee are reproduced from Ref.~\cite{Abada:2019zxq}.
}
\label{tab:detectors}
\end{center}
\end{table}

The individual decay probability of the neutralinos inside the fiducial volume of detectors is estimated with the following formulas:
\begin{eqnarray}
P[(\tilde{\chi}_1^0)_i \text{ in f.v.}] &=& e^{-L_i/\lambda_i^z} \cdot (1-e^{-L'_i/\lambda^z_i}),\\
L_i &\equiv & \text{min}(L_d,|R_I/\tan{\theta_i}|),\nonumber\\
L'_i &\equiv & \text{min}(L_d, |R_O/\tan{\theta_i}|)-L_i, \nonumber\\
\lambda_i^z&=& \beta_i^z \, \gamma_i \, c\,  \hbar / \Gamma_{\text{tot}}(\tilde{\chi}_1^0),
\label{eqn:individualdecprob}
\end{eqnarray}
where $\theta_i$ is the polar angle of an individual neutralino $(\tilde{\chi}_1^0)_i$, $\Gamma_{\text{tot}}(\tilde{\chi}_1^0)$ is the total decay width of the neutralino, $\beta^z_i$ is the velocity of $(\tilde{\chi}_1^0)_i$ along the beam axis, $\gamma_i$ is its Lorentz boost factor, and $c$ and $\hbar$ are the speed of light and the reduced Planck constant, respectively.

One potential issue regarding the background is the semi-leptonic decays of neutral SM mesons produced from $Z-$decays.
For the signal with light $\neutralino1$ (e.g. $m_{\neutralino1} \sim$ 1 GeV), the main sources of the background events are $D^0$ mesons decaying into a $K^\pm$, a charged lepton and a corresponding neutrino.
$D^0$ mesons mainly arise from either direct hadronization ($Z\rightarrow c\bar{c}$ process) or the decay of $B^\pm$ with a decay branching ratio of $\sim 5\%$.
By MC simulation with Pythia 8, we find that the typical decay distance\footnote{The decay distance is defined to be the distance from the IP to the displaced decay vertex.} of the $D^0$ mesons arising from these two sources
from the IP 
is in the range of 1-10 mm, as shown in Fig.~\ref{fig:decaydistanceD0}.
Since most $D^0$ mesons decay before travelling several centimeters,
these background events can be rejected by imposing a minimal radius requirement of the displaced vertex at $\sim$ several cm (e.g. 10 cm).

A rough conservative estimate can be made to justify this claim.
For the $D^0\bar{D}^0$ events stemming from $Z-$boson decays at a lepton collider, 
If one of them decays into $K^\pm e^\mp$ plus missing energy inside the inner detector with a decay branching ratio of $\sim 10\%$, and the other undergoes a ``0-prong'' decay with a decay branching ratio $\sim 15\%$, the signal of only one neutralino decaying in the inner detector can be faked.
Through MC simulation of 1 million events, we find that the mean boosted decay length of the $D^0$ mesons 
is $\sim$ 1 mm.
Therefore, assuming there are in total $10^{12}$ $D^0\bar{D}^0$ events, one can estimate the number of the events of the above mentioned type which remain outside a radius of $d^{D^0}_\text{decay}$ as $10^{12}\times(10\% \times 15\% \cdot 2)\times e^{(-d^{D^0}_\text{decay}/(1\text{ mm}))}$.
For one event of such kind, we obtain $d^{D^0}_\text{decay}\sim 2.4$ cm.
As a result, a minimal radius requirement of approximately $10$ cm should be able to render the selection efficiency of the semi-leptonic meson background down to below $10^{-12}$.
Note that this reflects again one advantage of lepton colliders running at the $Z-$pole mode, as at the LHC or a more energetic lepton collider, the mesons would be more strongly boosted, easily decaying with a larger distance (e.g. $\mathcal{O}(10)$ cm) and faking as the signal.

As for heavier $\neutralino1$, large ${\lambda'_{112}} / {m^2_{\tilde{f}}}$ values could lead to small $c \tau$, comparable to that of the mesons.
In such cases, according to Ref.~\cite{ATLAS:2019ems},  since the leptons from the meson decays are most often produced with nearby energy depositions from hadronic activities, such background processes can be rejected by requiring one well-isolated electron from nearby inner detector tracks and calorimeter deposits.
Besides, the majority of signals in the considered process will have an obvious time delay which may be exploited to suppress background as well~\cite{Liu:2018wte}.
Furthermore, the fast electronics and advanced analysis technology during the upgrades at future experiments would allow for a sophisticated object identification method, acting as a LLP tagger,  to veto the SM mesons.
Therefore, many strategies can be adopted to make such background under control. 
However, a realistic estimation of the background rejection efficiencies relies on the detailed information of the detector performance. Since the detector designs are still under development, we leave it for future studies.
For simplicity, in this study, we assume 100\% detector efficiency with no background event, and consider 3 signal events are sufficient for discovery of a long-lived neutralino.

Before we present sensitivity estimates of long-lived neutralinos, we first present the average decay probabilities for 1 GeV neutralinos at future lepton colliders and compare them with that of AL3X and MATHUSLA given in Ref.~\cite{Dercks:2018wum} for the same physics process $Z\rightarrow \tilde{\chi}_1^0\tilde{\chi}_1^0$ and neutralino mass. 
The average decay probability in the fiducial volume $\left\langle P[\tilde{\chi}_1^0 \text{ in f.v.}]\right\rangle$ is also known as fiducial efficiency, following the convention used in Refs.~\cite{Dercks:2018wum,Gligorov:2018vkc}.
We denote the fiducial efficiency for neutralinos pair produced from $Z-$decays as $\epsilon_{\text{fid}}^{Z \to \tilde{\chi}_1^0 \tilde{\chi}_1^0}$ and show its values at various experiments in Table~\ref{tab:fiducialefficiencies}.

\begin{table}[h]
\begin{center}
\begin{tabular}{c|c|c|c}
\hline
\hline
 & CEPC & FCC-ee CLD & FCC-ee IDEA \\
$\epsilon_{\text{fid}}^{Z \to \tilde{\chi}_1^0 \tilde{\chi}_1^0} \cdot c \tau$ [m]  
 & $4.78 \times 10^{-2}$ & $5.16 \times 10^{-2}$   &  $4.83\times 10^{-2}$  \\
\hline
 & AL3X & MATHUSLA & \\
$\epsilon_{\text{fid}}^{Z \to \tilde{\chi}_1^0 \tilde{\chi}_1^0} \cdot c \tau$ [m]
& $1.6\times 10^{-2}$ &  $8.0 \times 10^{-4}$ &   \\
\hline
\hline  
\end{tabular}
\caption{List of the fiducial efficiencies $\epsilon_{\text{fid}}^{Z \to \tilde{\chi}_1^0 \tilde{\chi}_1^0}$ multiplied by the proper decay length $c \tau$ in the unit of meter
for neutralinos of mass 1 GeV pair produced from $Z-$boson decays in the AL3X, MATHUSLA,  CEPC and FCC-ee detectors, for the boosted decay length much larger than the distance between the detector and the IP. The numbers for the cases of AL3X and MATHUSLA are reproduced from Ref.~\cite{Dercks:2018wum}.
}
\label{tab:fiducialefficiencies}
\end{center}
\end{table}

We work in the limit that the boosted decay length $\beta\gamma c \tau$ of the neutralino is much larger than the distance from the IP to the detector, such that we are allowed to present $\epsilon_{\text{fid}}^{Z \to \tilde{\chi}_1^0 \tilde{\chi}_1^0}$ with a linear dependence on $c \tau$, though in the calculation we use the exact formula.
The typical values of $\beta\gamma$ of 1 GeV neutralinos produced from $Z-$bosons at the CEPC and the FCC-ee are $\sim 45$.
Therefore, our results are legitimate for $c\tau \geq 1$ m for both the CEPC and the FCC-ee. 
We find that for large decay length of neutralinos in the lab frame, the detectors of the future lepton colliders show a similar fiducial efficiency that is larger than that of AL3X and MATHUSLA in this benchmark scenario.
This better efficiency is partly due to the almost full coverage of polar and azimuthal angle of the detectors at lepton colliders, and partly due to the fact that the $Z-$pole center-of-mass energy leads to the produced $Z-$bosons almost stationary and hence their decay products, i.e.\,the neutralinos, less boosted in the forward direction.
The similarity of the fiducial efficiency between the CEPC baseline detector and the FCC-ee CLD/IDEA, is consistent with the closeness of their geometrical parameters listed in Table.~\ref{tab:detectors}.

\section{Numerical Results}
\label{sec:results}

In this section, we present our numerical results.
Following the choice made in Refs.~\cite{Helo:2018qej,Dercks:2018wum}, we consider two benchmark values for BR$(Z\rightarrow \tilde{\chi}_1^0\tilde{\chi}_1^0)$\footnote{The numbers of BR$(Z\rightarrow \tilde{\chi}_1^0\tilde{\chi}_1^0)$ refer to the case when $m_{\tilde{\chi}_1^0}\ll m_Z$.
For larger neutralino masses, we have taken into account the phase space suppression effect in our evaluation.}: the experimental upper limit $10^{-3}$ and a more conservative choice $10^{-5}$.
We choose to require $\lambda'_{112}$ $L_1 \cdot Q_1 \bar{D}_2$ as the only nonvanishing RPV operator, which leads to the lightest neutralino decays to SM particles via a sfermion exchange.

\begin{figure}[h]
\centering
\includegraphics[width=0.48\textwidth]{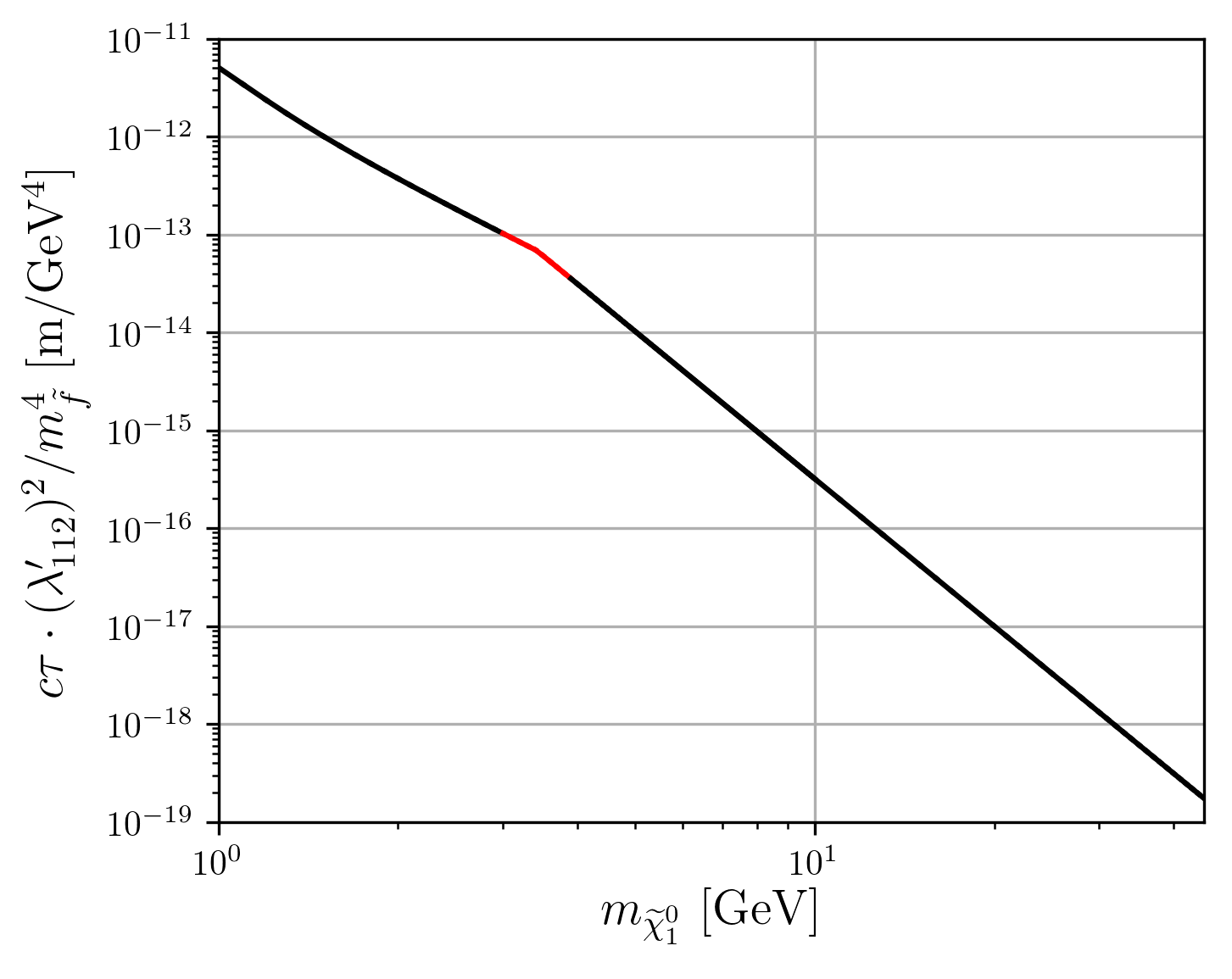}
  \caption{ The proper decay length of $\tilde{\chi}_1^0$ times $\lambda_{112}^{\prime 2} / m^4_{\tilde{f}}$ vs. $m_{\tilde{\chi}_1^0}$.
  }
\label{fig:decaylength}
\end{figure}

The decay mode depends on the neutralino mass.
For $m_{\tilde{\chi}_1^0}$ of $\mathcal{O}(\text{GeV})$ mass, the hadronization effects are important and 
\begin{eqnarray}
 \tilde{\chi}_1^0 \rightarrow 
 \begin{cases}
 (K^0_L, K^0_S,K^*)+(\nu_e,\bar{\nu}_e) \text{, invisible mode,}\\
 (K^\pm,K^{*\pm})+e^\mp \text{, visible mode,}
 \end{cases}
 \label{eqn:decaymodes}
\end{eqnarray}
while heavier neutralinos would decay to two jets and an electron/missing energy (visible/invisible mode).
For neutralinos undergoing the two-body decay modes given in Eq.~\eqref{eqn:decaymodes} we calculate the neutralino decay width with the formulas given in Ref.~\cite{deVries:2015mfw}.
In order to calculate three-body decay widths for heavier neutralinos, we use the three-body decay $(\tilde{\chi}_1^0\rightarrow e^{\mp}/\nu_e+jj)$ results given by SPheno 4.0.3 \cite{Porod:2003um,Porod:2011nf} and parameterize the lightest neutralino proper decay length as:
\begin{eqnarray}
c\tau=(3.2~\text{m})\bigg(\frac{m_{\tilde{f}}}{1~\text{TeV}}\bigg)^4\,\bigg(\frac{10~\text{GeV}}{m_{\tilde{\chi}_1^0}}\bigg)^5\, \bigg(\frac{0.01}{\lambda'_{112}} \bigg)^2,
\label{eqn:decayLength}
\end{eqnarray} 
where $m_{\tilde{f}}$ is the relevant sfermion mass and we have normalized $m_{\tilde{f}}$, $m_{\tilde{\chi}_1^0}$ and $\lambda'_{112}$ to their typical values.
We find that the two-body and three-body decay width formulas converge at $m_{\tilde{\chi}_1^0}\sim 3.5$ GeV.
Therefore, for the numerical calculation of the decay width of $\tilde{\chi}_1^0$, we simply take 3.5 GeV as the threshold between the two-body and three-body decays.
As the decay width of $\tilde{\chi}_1^0$ is proportional to $\lambda_{112}^{\prime 2} / m^4_{\tilde{f}}$ for the whole relevant mass range, we present a plot in Fig.~\ref{fig:decaylength} for the neutralino proper decay length $c\tau$ multiplied by $\lambda_{112}^{\prime 2} / m^4_{\tilde{f}}$ as a function of the neutralino mass $m_{\tilde{\chi}_1^0}$.
In Fig.~\ref{fig:decaylength} the transition mass range between 2-body and 3-body decays is highlighted with red color, emphasizing the artificial nature of the kink at 3.5 GeV.
For a typical value of $\lambda_{112}^{\prime 2} / m^4_{\tilde{f}}=10^{-16}$ GeV$^{-4}$ for instance, $c\tau \approx 5\times 10^4~\text{m}$ for $m_{\tilde{\chi}_1^0}=1$ GeV and $c\tau \approx 0.1~\text{m}$ for $m_{\tilde{\chi}_1^0}=20$ GeV.

In this study, when calculating $N_{\tilde{\chi}_1^0}^{\text{obs}}$ with Eq.~\eqref{eqn:obsneu1number} we consider two cases for the $\tilde{\chi}_1^0$ decays: (i) all the final states of decays can be identified so that $\text{BR}(\tilde{\chi}_1^0 \rightarrow \text{ final state}) = 100\%$; (ii) only the visible/charged final states of decays can be identified so that $\text{BR}(\tilde{\chi}_1^0 \rightarrow \text{ final state}) = \text{BR}(\tilde{\chi}_1^0 \rightarrow \text{ visible mode only})$.
Since the visible/charged products are usually easier to be reconstructed in the detectors, the latter is more conservative.

\begin{figure}[t]
\centering
\includegraphics[width=0.48\textwidth]{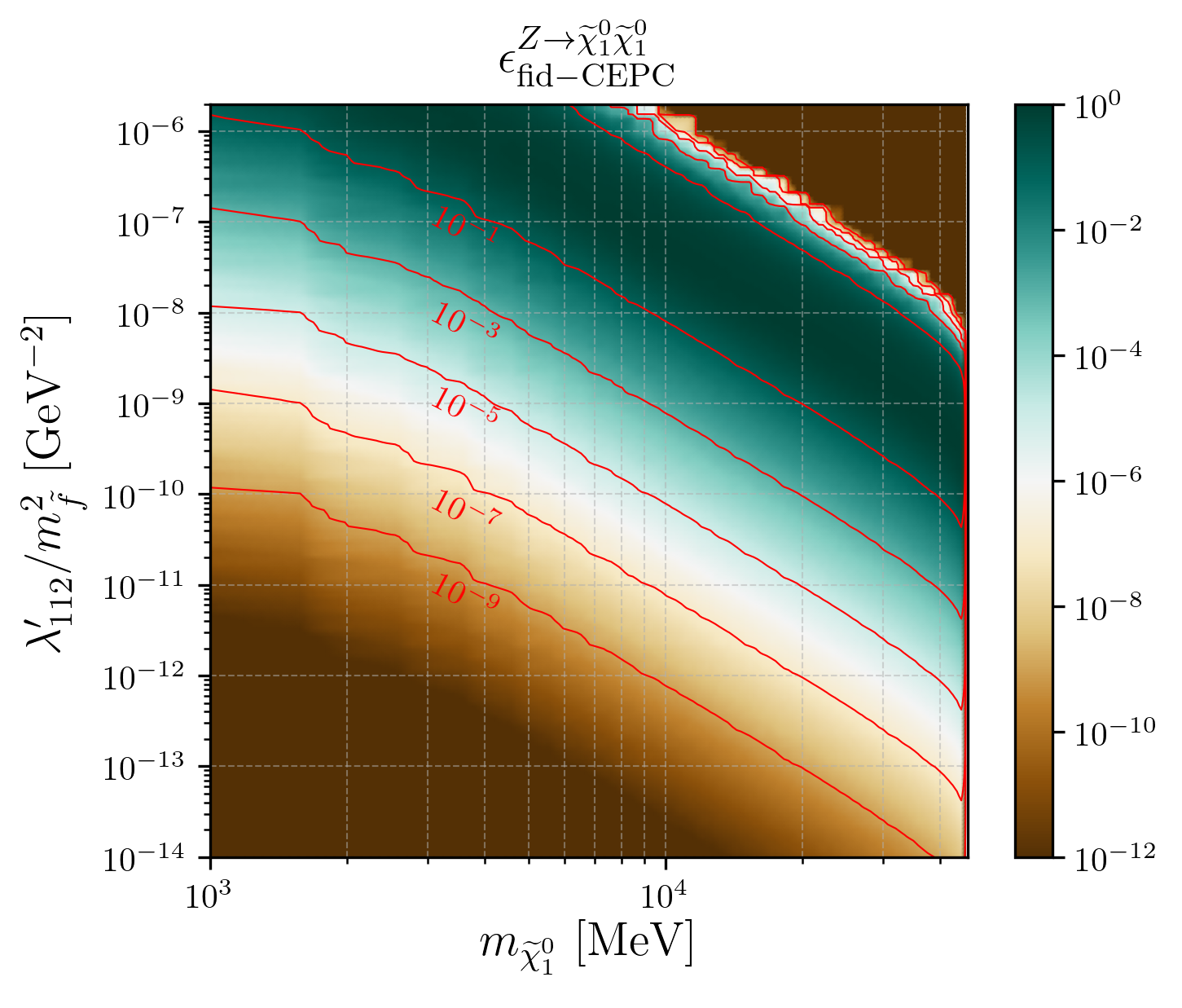}
  \caption{ Fiducial efficiency $\epsilon_{\text{fid}}^{Z\rightarrow \tilde{\chi}_1^0 \tilde{\chi}_1^0}$ of the CEPC baseline detector as a function of $m_{\tilde{\chi}_1^0}$ and $\lambda'_{112} / m^2_{\tilde{f}}$. 
The red contour curves correspond to $\epsilon_{\text{fid}}^{Z\rightarrow \tilde{\chi}_1^0 \tilde{\chi}_1^0}=10^{-1},10^{-3}, 10^{-5}, 10^{-7}, 10^{-9}$, respectively. }
\label{fig:fideff}
\end{figure}

\begin{figure}[t]
\centering
\includegraphics[width=0.96\columnwidth]{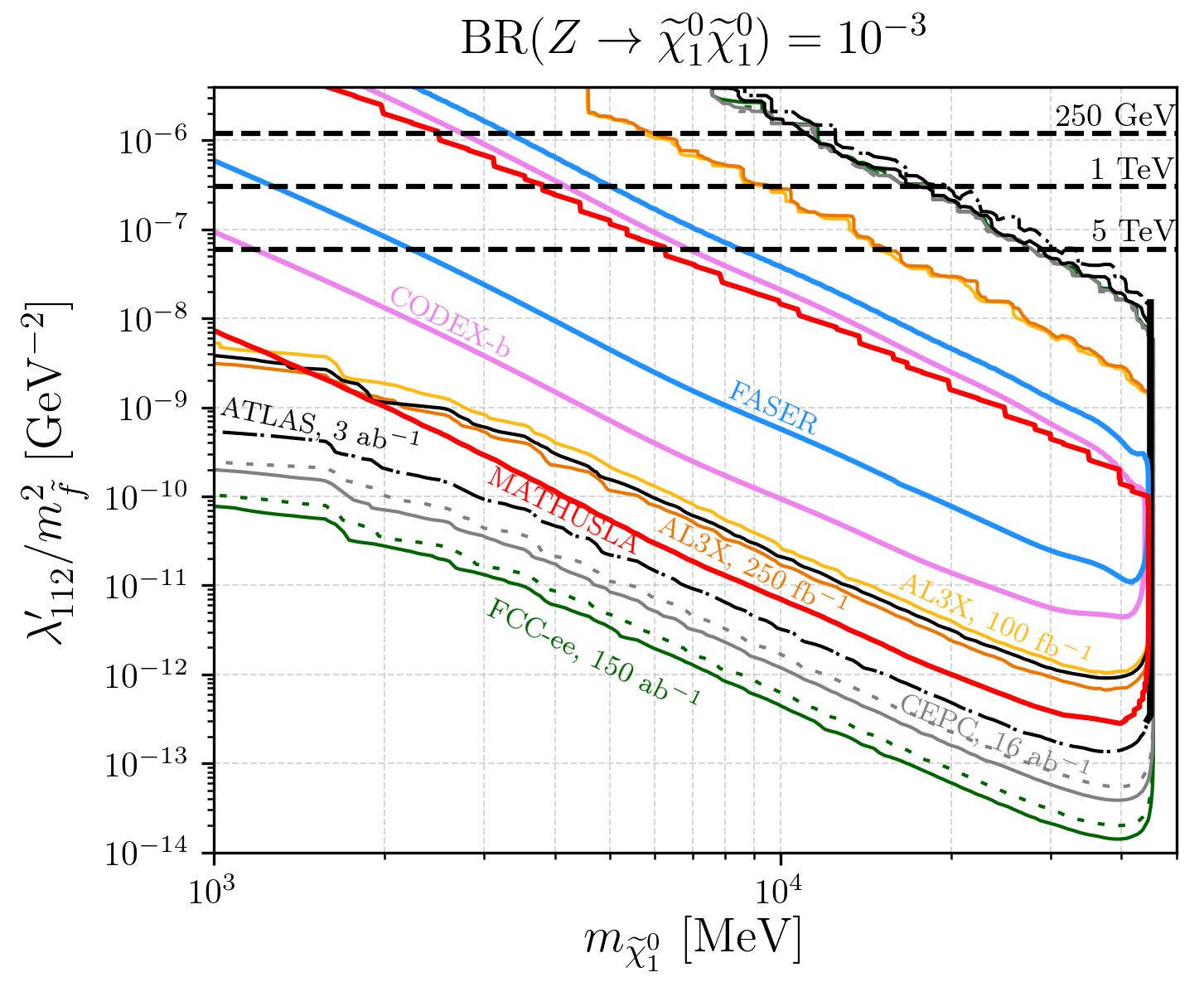}
\includegraphics[width=0.96\columnwidth]{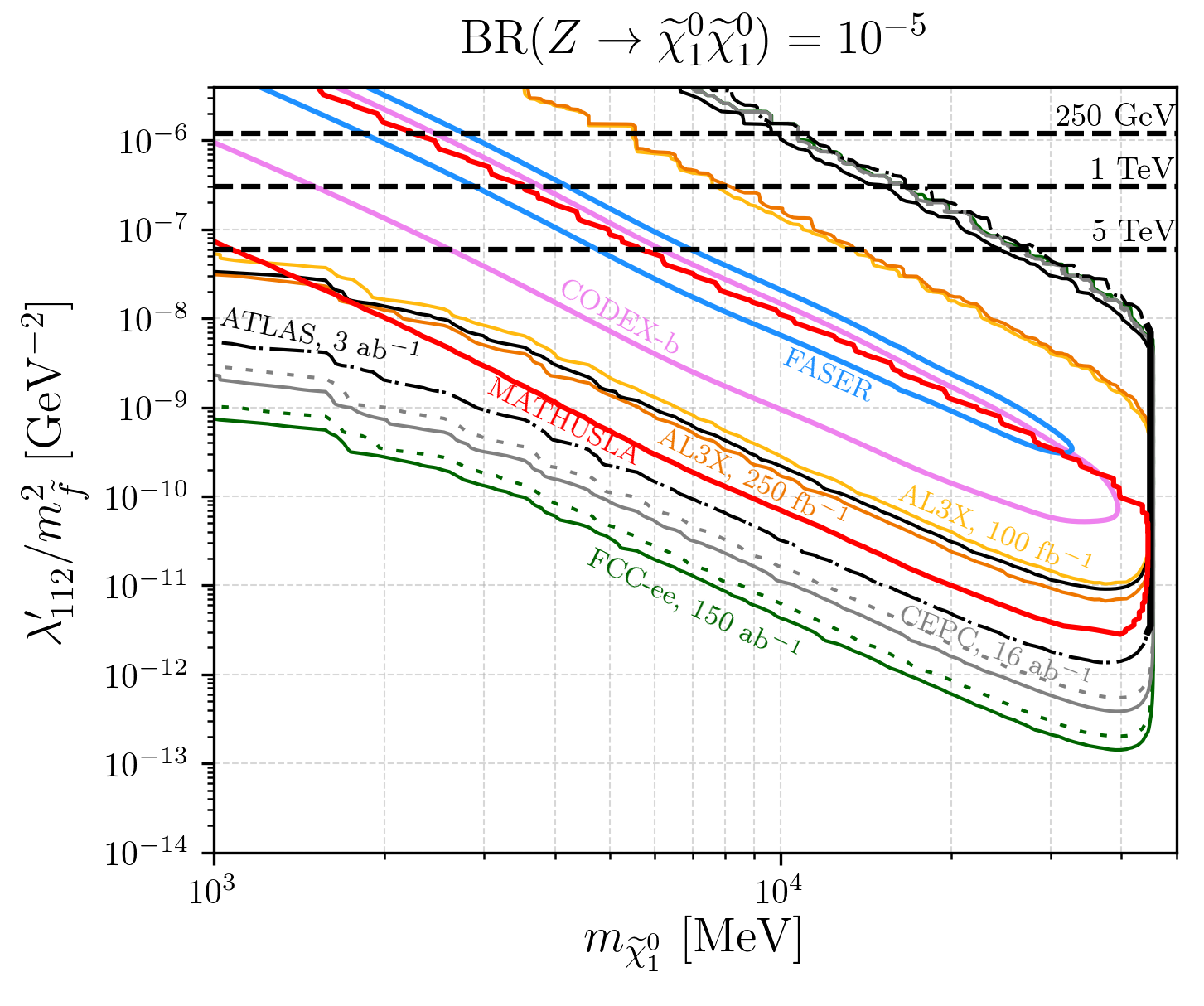}
\caption{The sensitivity estimate of the CEPC (grey) and the FCC-ee (green) presented in the 2D plane of $\lambda'_{112}/m_{\tilde f}^2$ vs. $m_{\tilde{\chi}_1^0}$ for two different benchmark values of BR($Z\rightarrow \tilde{\chi}_1^0 \tilde{\chi}_1^0$), respectively.
The solid contour curves correspond to three decay events in the fiducial volume when considering all decay modes of $\tilde{\chi}_1^0$, while the dashed lines include only visible/charged decay modes ($K^{(*)\pm} e^\mp$, $e^-us$ or $e^+\bar{u}\bar{s}$).
The estimates for experiments at the LHC: AL3X, CODEX-b, FASER and MATHUSLA, are reproduced from Refs.~\cite{Helo:2018qej,Dercks:2018wum}.
The ATLAS results correspond to HL-LHC for $\sqrt{s}=14$ TeV and 3 $\iab$ integrated luminosity.
The black horizontal dashed lines correspond to the current RPV bounds on the single coupling $\lam^\prime_{112}$ \cite{Kao:2009fg} for three different degenerate sfermion masses $m_{\tilde{f}}=250$ GeV, 1 TeV, and 5 TeV as labelled.
}
\label{fig:zneuneuplots}
\end{figure}

In Fig.~\ref{fig:fideff}, we present the distribution of the fiducial efficiency $\epsilon_{\text{fid}}^{Z\rightarrow \tilde{\chi}_1^0 \tilde{\chi}_1^0}$ of the CEPC baseline detector in the $(m_{\tilde{\chi}_1^0},\lambda'_{112} / m^2_{\tilde{f}})$ plane. 
We overlap the plot with red isocurves for a set of $\epsilon_{\text{fid}}^{Z\rightarrow \tilde{\chi}_1^0 \tilde{\chi}_1^0}$ values.
We find that in certain parts of the parameter space the fiducial efficiency can be close to $1$.
The corresponding plots for the FCC-ee CLD and IDEA detectors are very similar and we hence refrain from showing them here.

In Fig.~\ref{fig:zneuneuplots}, 
we present two plots of 3-signal-event\footnote{For a background-free study, 3 signal events correspond to 95\% confidence level (C.L.) limits.} contour curves in the $\lambda'_{112} / m^2_{\tilde{f}}$ vs. $m_{\tilde{\chi}_1^0}$ plane for the two benchmark values of BR$(Z\rightarrow \tilde{\chi}_1^0\tilde{\chi}_1^0)$.
For simplicity, we assume all the sfermions are degenerate in mass during the evaluation.
We show with three hashed horizontal lines the current upper limit on $\lambda'_{112}$ for three benchmark sfermion mass values: 250 GeV, 1 TeV and 5 TeV, extracted from Ref.~\cite{Kao:2009fg}:
\begin{eqnarray}
\lambda'_{112}<0.6 \,\, \frac{m_{\tilde{s}_R}}{2 \text{ TeV}}.
\end{eqnarray}
As for the experimentally excluded region in the parameter space, one may recast the DELPHI search for heavy neutral leptons in $Z$ decays \cite{Abreu:1996pa}, as was done in Ref.~\cite{Helo:2018qej} which one may see for reference.
These limits roughly exclude the parameter region above the lower half of the FASER contour curve.
We refrain from showing them in Fig.~\ref{fig:zneuneuplots} in order to keep the plots legible.

Since various detectors of the $e^+ e^-$ colliders possess a similar fiducial efficiency, we show isocurves of the CEPC baseline detector and the FCC-ee IDEA detector only.
The green (grey) solid lines show the limits at the FCC-ee (CEPC) with the IDEA (baseline) detector design and 150 (16) $\iab$ integrated luminosity when including all decay modes of the lightest neutralinos, while the dashed curves are limits when including only the visible/charged decay modes.
We overlap the plots with estimates from other experiments at the LHC: AL3X, CODEX-b, FASER and MATHUSLA, extracted from Refs.~\cite{Helo:2018qej,Dercks:2018wum}.

To complete the physics picture, it is also necessary to compare the CEPC/FCC-ee limits to those at the ATLAS or CMS detector at the HL-LHC.
We estimate sensitivities for our signal scenario at HL-LHC with $\sqrt{s}=14$ TeV and 3 $\iab$ integrated luminosity.
Following the procedure in Ref.~\cite{Helo:2018qej}, the number of $Z-$boson produced at HL-LHC is estimated as $N_Z \approx 1.8\times10^{11}$.
We choose to focus on the ATLAS detector for the HL-LHC.
Its shape is the same as the CEPC baseline detector and we extract its geometry sizes from Ref.~\cite{Aaboud:2017pjd}: $R_I=$ 0.0282 m, $R_O=$ 1.1 m, and $L_d =$ 3.1 m (cf. Fig.~\ref{fig:detector-sketch}).
For neutralinos of mass 1 GeV pair produced from $Z-$boson decays at the HL-LHC, the fiducial efficiency multiplied by the proper decay length is calculated to be $\epsilon_{\text{fid}}^{Z \to \tilde{\chi}_1^0 \tilde{\chi}_1^0} \cdot c \tau = 2.39\times10^{-2}$ m in the large boosted decay length limit, which is slightly smaller than these values at the CEPC/FCC-ee (cf. Table~\ref{tab:fiducialefficiencies}).
Similar to the study at the CEPC/FCC-ee, assuming 100\% detector efficiency and background free environment, we show the 3-signal-event sensitivities at ATLAS as dot-dashed black curves in Fig.~\ref{fig:zneuneuplots}.

It is worth noting that because of the large background at the LHC, the sensitivities at ATLAS could be reduced significantly with the background taken into account.
To quantify this reduction from the SM background,
since there is no LHC experimental searches related to $Z-$bosons decaying to a pair of LLPs,
we extrapolate the number of background events from the ATLAS study~\cite{Aad:2019xav}.
This search, with center-of-mass energy at 13 TeV and integrated luminosity of 33 fb$^{-1}$, considered a process of the SM Higgs decaying to a pair of long-lived light scalars which further decay to a pair of jets each.
It is hence featured with similar kinematics as this study.
In Ref.~\cite{Aad:2019xav}, when requiring at least one reconstructed displaced vertex in the inner detector which passes all relevant selection requirements in the background region, the number of background events is 45
\footnote{
We do not use the number of background events in the signal region, as its selection requirements are more strict than those in the background region.
Using this number would lead to underestimation of the background events.
In fact, even for the background region, some requirements like applying a veto on the Muon RoI Cluster trigger are imposed, leading to underestimation of the background events.
}.
Re-scaling this number by the integrated luminosity ratio ( $3000~\ifb / 33~\ifb \simeq 91$), and assuming the kinemtics and production cross section do not change too much between 13 and 14 TeV, the number of background events is expected to be 4091 at the HL-LHC.
Since the final state is similar, the signal considered in this study would hence also suffer from this amount of background.
In order to calculate the number of signal events corresponding to 95\% C.L., we require the signal significance $Z^{\text{sig.}}\simeq 2$
\footnote{
The signal significance is defined as $Z^{\text{sig.}} \equiv \sqrt{2\cdot ((s+b)\cdot \text{ln}(1+s/b)-s)}$, with $s$ and $b$ here denoting the expected number of signal and background events, respectively.
}.
The corresponding signal events are calculated to be 129.
Consequently, we include a solid black iso-curve in Fig.~\ref{fig:zneuneuplots}, representing the ATLAS limits with the background taken into account.
We emphasize that this rough estimate is rather for illustrating the effect of the background processes at the LHC than for providing a concrete conclusion.
Detailed data simulation and optimized analysis are necessary for more realistic limits of our physics scenario at the LHC, which is beyond the scope of this work and is left for future studies.

We observe that all detectors may have a sensitivity reach in $\lambda'_{112} / m^2_{\tilde{f}}$ for the whole range of the neutralino mass orders of magnitude smaller than the current RPV upper bounds. 
At the LHC, after taking into account the large irreducible background source, the ATLAS experiment shows similar lower sensitivity reach as AL3X, while MATHUSLA clearly outperforms both ATLAS and other proposed far-detector experiments.
The future lepton colliders enclose all the sensitive parameter space covered by the future LHC experiments. 
This is mainly because a terascale number of $Z-$bosons can be produced at the future lepton colliders, and also because the lepton colliders offer a relatively clean environment for such LLP searches.

At the small-value regime, when $m_{\tilde{\chi}_1^0} \sim 40$ GeV and BR$(Z\rightarrow \tilde{\chi}_1^0\tilde{\chi}_1^0) = 10^{-3}$, the FCC-ee can reach as low as $1.5\times 10^{-14}$ GeV$^{-2}$ in $\lambda'_{112} / m^2_{\tilde{f}}$ with 150 $\iab$ luminosity, while the CEPC reaches $3.9\times 10^{-14}$ GeV$^{-2}$ with luminosity of 16 $\iab$.
Since their fiducial efficiencies are similar, this difference in sensitivities is almost fully due to the difference in luminosities.
For BR$(Z\rightarrow \tilde{\chi}_1^0\tilde{\chi}_1^0) = 10^{-5}$, the FCC-ee's lower reach in $\lambda'_{112} / m^2_{\tilde{f}}$ can still be down to $1.5\times 10^{-13}$ GeV$^{-2}$, and the upper reach of the FCC-ee and the CEPC does not change much compared to the larger BR$(Z\rightarrow \tilde{\chi}_1^0\tilde{\chi}_1^0)$ case.

Note in both plots, the lower bound of the CEPC/FCC-ee dashed curves, which indicates the limits when including only visible/charged decay modes of $\tilde{\chi}_1^0$,  is only slightly worse than the solid isocurves.
This is because in almost the whole kinematically allowed mass range (except when $m_{K^{\pm}}<m_{\tilde{\chi}_1^0}<m_{K^0_{L/S}}$ ), the visible decay branching ratio of the $\tilde{\chi}_1^0$ is approximately 0.5.

\section{Conclusions}
\newcommand{\nicefrac}[2]{#1/#2}
\label{sec:conclusions}

In this study, we estimate the sensitivity reaches of future high-luminosity $Z-$factories when detecting long-lived particles produced from $Z-$boson decays.
We discuss the general advantages of the CEPC and FCC-ee running at the $Z-$pole mode when detecting such $Z$-coupled LLPs compared to the LHC.
To demonstrate the limits, we consider one particular benchmark scenario: the long-lived lightest neutralinos pair produced from on-shell $Z-$boson decays ($Z\rightarrow \tilde{\chi}_1^0\tilde{\chi}_1^0$) in the RPV-SUSY.
The results are shown in Fig.~\ref{fig:zneuneuplots}.
The two plots in the $\lambda'_{112} / m^2_{\tilde{f}}$ vs.~$m_{\tilde{\chi}_1^0}$ plane correspond to two benchmark values of BR$(Z\rightarrow \tilde{\chi}_1^0 \tilde{\chi}_1^0)$: the experimental upper constraint $10^{-3}$ and a more conservative choice $10^{-5}$.
We find that the Z-pole running mode at future lepton colliders has a sensitivity reach in $\lambda'_{112} / m^2_{\tilde{f}}$ orders of magnitude smaller than the current RPV upper limits for the mass range $1$ GeV $\lsim m_{\tilde{\chi}_1^0} \lsim m_Z/2$. 
When $m_{\tilde{\chi}_1^0} \sim 40$ GeV and BR$(Z\rightarrow \tilde{\chi}_1^0\tilde{\chi}_1^0) = 10^{-3}$ , FCC-ee can reach as low as $1.5\times 10^{-14}$ GeV$^{-2}$ in $\lambda'_{112} / m^2_{\tilde{f}}$ with 150 $\iab$ luminosity, while CEPC reaches $3.9\times 10^{-14}$ GeV$^{-2}$ with luminosity of 16 $\iab$.
The FCC-ee has stronger sensitivity reaches mainly by virtue of its larger integrated luminosity than that of the CEPC.
Moreover, lepton colliders could not only enclose all the sensitive parameter space covered by the HL-LHC ATLAS experiments and other proposed far-detector experiments at the LHC including AL3X, CODEX-b, FASER and MATHUSLA, but also extend both the upper and lower reaches in $\lambda'_{112} / m^2_{\tilde{f}}$ by more than one order of magnitude. 
This is mainly because compared with the HL-LHC, the future lepton colliders can produce many more $Z-$bosons, and the detector setups of lepton colliders also can have a slightly larger coverage of the solid angle.

Our results show that the unprecedentedly large number of $Z-$bosons expected to be produced at the future $Z-$factories may serve as a very sensitive probe of exotic decays of $Z-$bosons into LLPs.
Our work on the lightest neutralinos in the context of the RPV-SUSY complements the other studies in the literature on rare $Z-$decays.
Serving as one benchmark example, this particular physics case exemplifies the unique advantages of the next-generation $e^+ e^- -$colliders for probing $Z$-coupled LLPs, compared to the present and future LHC experiments.

\begin{acknowledgments}

We thank Herbi Dreiner, Oliver Fischer, Manqi Ruan, Torbj\"{o}rn Sj\"{o}strand, Yuexin Wang, and Rui Zhang for useful discussions.
Special thanks go to the referees for helping us to improve the quality of this work.
K.W. acknowledges supports from the Excellent Young Talents Program of the Wuhan University of Technology, the CEPC theory grant (2019-2020) of IHEP, CAS and the National Natural Science Foundation of China under grant no.~11905162.
Z.S.W. is supported by the Sino-German DFG grant SFB CRC 110 ``Symmetries and the Emergence of Structure in QCD", the Ministry of Science, ICT \& Future Planning of Korea, the Pohang City Government, and the Gyeongsangbuk-do Provincial Government through the Young Scientist Training Asia-Pacific Economic Cooperation program of APCTP.

\end{acknowledgments}

\appendix

\section{Kinematic Distributions of Decay Products}
\label{sec:triggers}

\begin{figure}[h]
	\centering
	\includegraphics[width=\columnwidth]{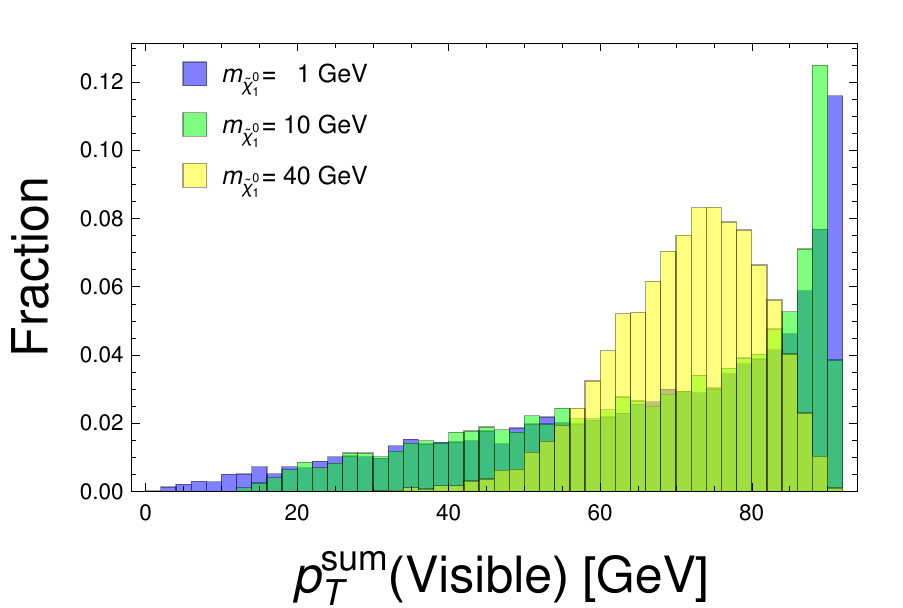}
	\includegraphics[width=\columnwidth]{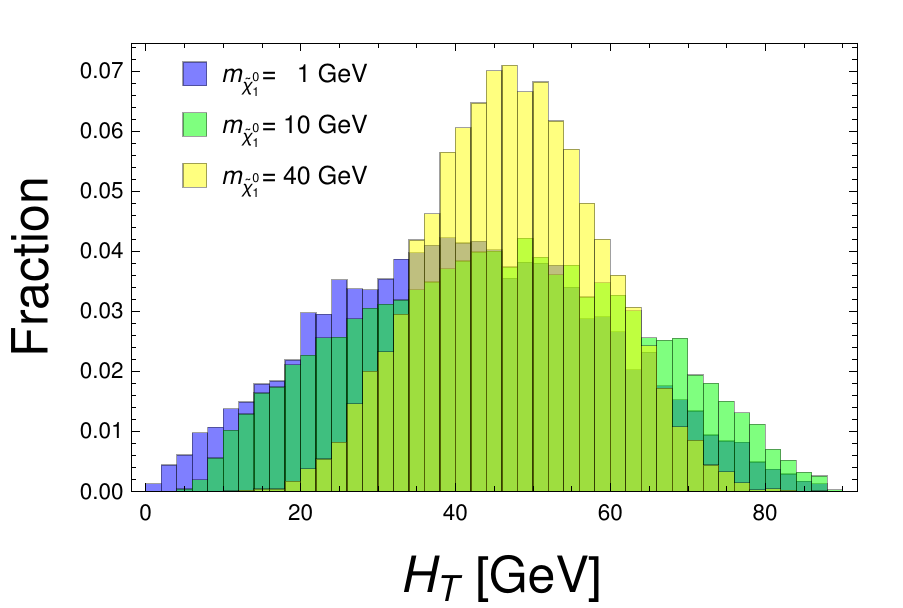}
   \caption{
	Kinematic distributions of decay products from two neutralinos for three benchmark $m_{\tilde{\chi}_1^0}$ values when both neutralinos decay inside the inner detector. 
	$p_T^{\rm sum}(\rm Visible)$ denotes the sum of transverse momentum $p_T$ of all visible decay products, while $H_T$ is the sum of $p_T$ of all hadronic decay products.
}
	\label{fig:trigger_distribution_1}
\end{figure}

\begin{figure}[h]
	\centering
	\includegraphics[width=\columnwidth]{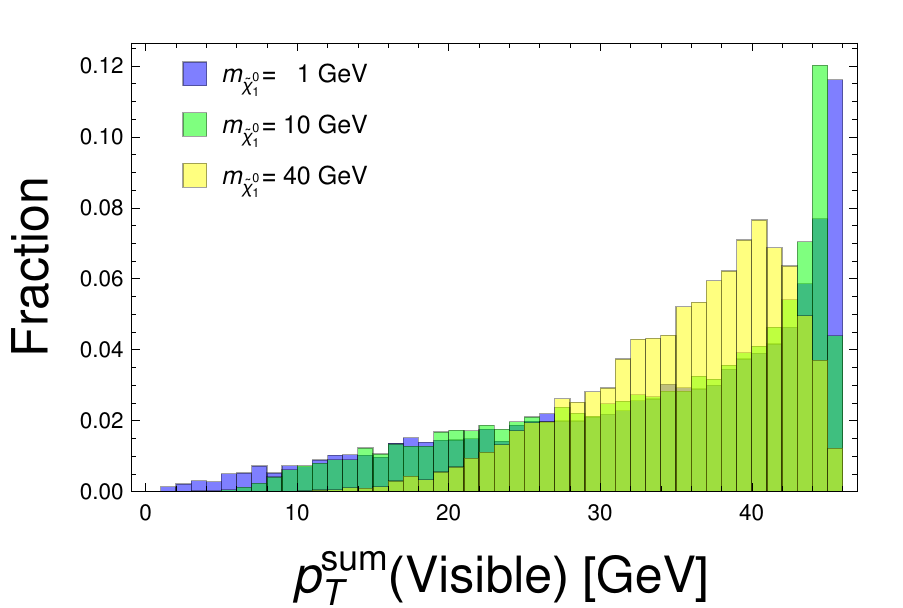}
	\caption{
	Kinematic distributions of decay products for three benchmark $m_{\tilde{\chi}_1^0}$ values when only one neutralino decays inside the inner detector. 
	$p_T^{\rm sum}(\rm Visible)$ is the sum of $p_T$ of all visible decay products from the decaying neutralino.
	}
	\label{fig:trigger_distribution_2}
\end{figure}

\section{Decay Distance of $D^0$ Mesons }
\label{sec:Dmeson}

\begin{figure}[H]
	\centering
	\includegraphics[width=\columnwidth]{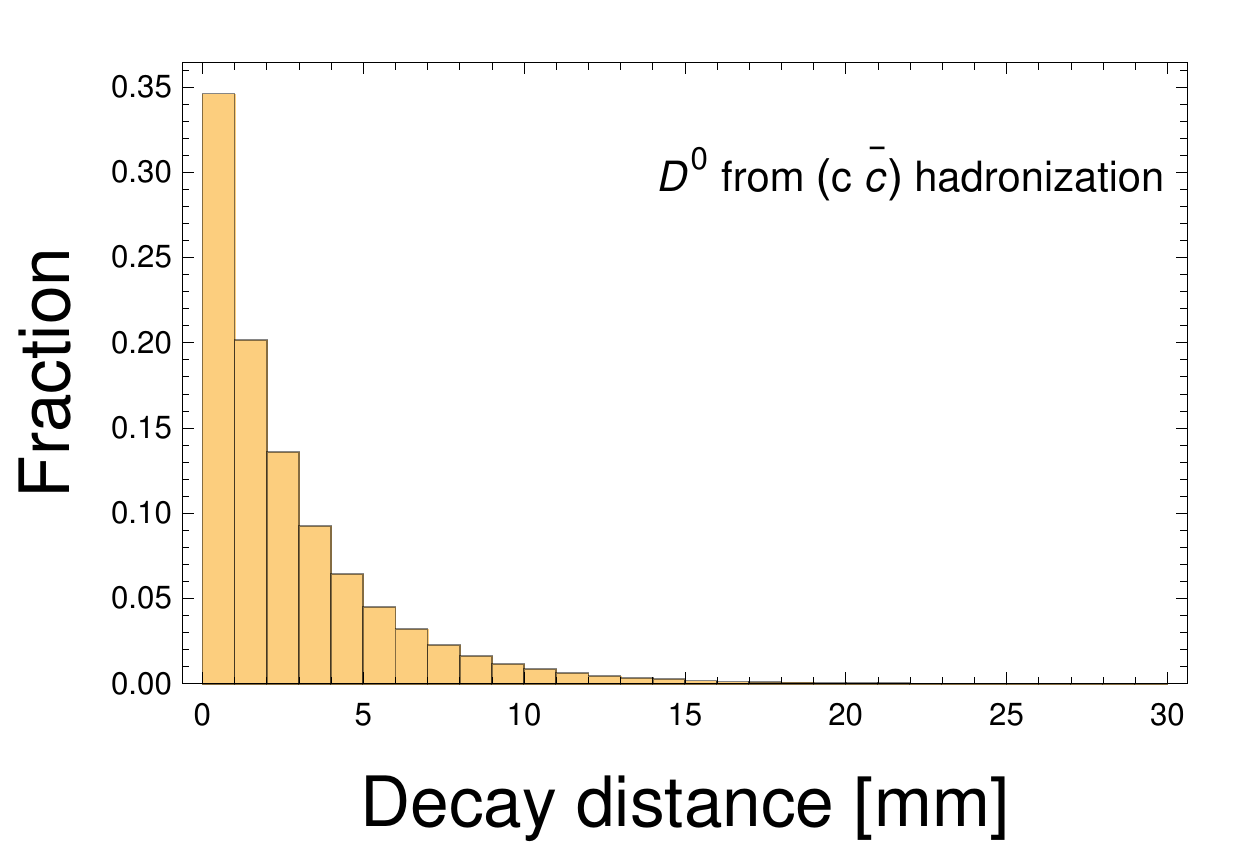}
	\includegraphics[width=\columnwidth]{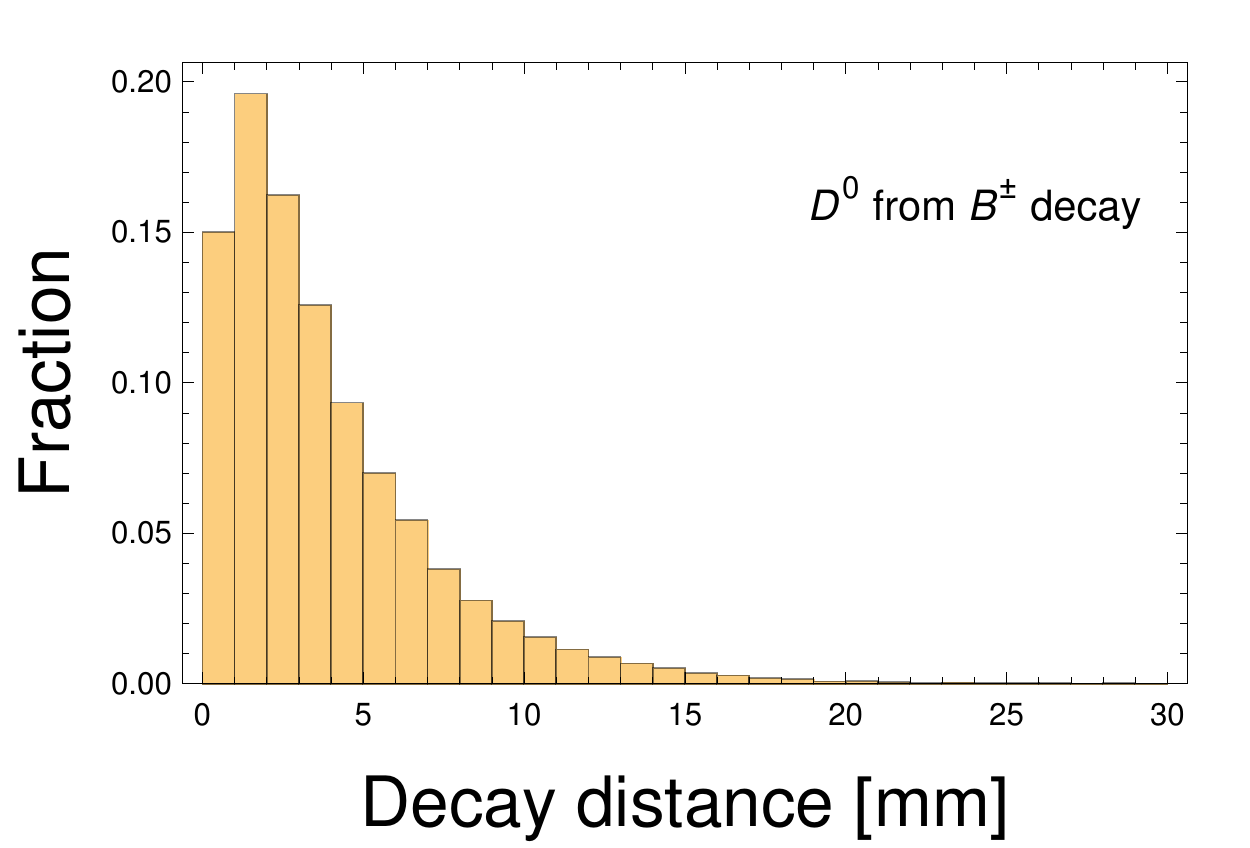}
	\caption{Distribution of the decay distance of $D^0$ mesons arising from direct hadronization ($c\bar{c}$) and $B^\pm$ decay. ``Decay distance'' is defined as the distance between the IP and the displaced decay vertex.	
	}
	\label{fig:decaydistanceD0}
\end{figure}

\bibliographystyle{h-physrev5}

\end{document}
\grid